\documentstyle[twoside,fleqn,espcrc2,epsf]{article}

\title{\vspace{-.5in}\hbox{}\hfill{\normalsize ANL-HEP-CP-96-55}\\
\hfill\\
QCD with chiral 4-fermion interactions ($\chi$QCD)%
\thanks{Talk presented by D.~K.~Sinclair at LATTICE'96, St.~Louis MO,
4--8 June, 1996.}}

\author{J~.B.~Kogut\address{Department of Physics, University of Illinois,
                            1110 West Green Street, Urbana, IL 61801}%
        \thanks{Supported in part by NSF grant NSF-PHY92-00148.}
        and
        D.~K.~Sinclair\address{HEP Division, Argonne National Laboratory,
                               9700, South Cass Avenue, Argonne, IL 60439}%
        \thanks{Supported by DOE contract W-31-109-ENG-38.}}
        
\begin{document}

\begin{abstract}
Lattice QCD with staggered quarks is augmented by the addition of a chiral
4-fermion interaction. The Dirac operator is now non-singular at $m_q=0$, 
decreasing the computing requirements for light quark simulations by at least
an order of magnitude. We present preliminary results from simulations at
finite and zero temperatures for $m_q=0$, with and without gauge fields.
\end{abstract}

\maketitle

\section{INTRODUCTION}

$\chi$QCD is QCD with a chiral 4-fermion interaction
(Nambu--Jona-Lasinio---Gross--Neveu)\cite{nambu-jona-lasinio,gross-neveu}. 
Since the 4-fermion term is an irrelevant operator, this theory should lie in
the same universality class as regular QCD. 

On the Lattice we use staggered quarks. For simplicity we consider a theory
where the 4-fermion operator has the 
$$
     U(1) \times U(1)    \subset     SU(N_f) \times SU(N_f)
$$
flavour symmetry generated by $(1,i\gamma_5\xi_5)$ \cite{4-fermi}. The
molecular dynamics Lagrangian for the lattice theory is 
\begin{eqnarray}
L & = & -\beta\sum_{\Box}[1-\frac{1}{3}{\rm Re}({\rm Tr}_{\Box} UUUU)]
        +\sum_s \dot{\psi}^{\dag} A^{\dag} A\dot{\psi}           \nonumber  \\
  &   & -\sum_{\tilde{s}}\frac{1}{8}N_f\gamma(\sigma^2+\pi^2)      
        +\frac{1}{2}\sum_{\tilde{s}}(\dot{\sigma}^2+\dot{\pi}^2) \nonumber  \\
  &   & +\frac{1}{2}\sum_l(\dot{\theta}_7^2+\dot{\theta}_8^2
        +\dot{\theta}_1^{\ast}\dot{\theta}_1
        +\dot{\theta}_2^{\ast}\dot{\theta}_2
        +\dot{\theta}_3^{\ast}\dot{\theta}_3)
\end{eqnarray}
where 
\begin{equation}
A = \not\!\! D + m_q + \frac{1}{16} \sum_i (\sigma_i+i\epsilon\pi_i)
\end{equation}
with $\epsilon=(-1)^{x+y+z+t}$. This describes 8 flavours. For $N_f$ which is
not a multiple of 8 we use ``noisy'' fermions \cite{noise} and multiply the
fermion kinetic term by $N_f/8$. 

The Dirac operator of regular lattice QCD becomes singular as $m_q \rightarrow
0$, whereas that for $\chi$QCD remains non-singular at $m_q=0$. Conjugate
gradient inversion of the regular QCD Dirac operator requires a number of
iterations which $\rightarrow \infty$ as $m_q \rightarrow 0$. Inversion of
$\chi$QCD Dirac operator requires a finite number of iterations even at
$m_q=0$. In practical terms, conjugate gradient inversion of the regular QCD
Dirac operator requires $\sim$10,000 iterations at the physical u and d quark
masses. Simulating $\chi$QCD at $m_q=0$, the worst we have found is $\sim 400$
iterations --- a saving of a factor of $\sim 12$ in CPU time! $\chi$QCD allows
us to work {\it at} the chiral limit, which is often useful.

All the simulations which I will now describe are being performed at $m_q=0$.
Simulations are being performed on the CRAY C-90 and CRAY J-90 at NERSC.

\section{$\chi$QCD AT FINITE TEMPERATURE}

We are studying the finite temperature behaviour of $\chi$QCD with $N_f=2$
and $m_q=0$, using the hybrid molecular dynamics algorithm with ``noisy''
quarks, as a function of $\beta$ and $\gamma$.

We are studying the deconfinement and chiral transitions --- not in general
coincident at finite $a$. The 2 transitions come together as $a \rightarrow 0$
and in general as $\gamma$ is increased. By measuring the critical behaviour of
such observables as $\langle\bar{\psi}\psi\rangle$, we hope to determine the
equation of state, where these transitions coincide. For this, being able to
work {\it at} the chiral limit is a boon.

Our present simulations are being performed at $N_t=4$, using $8^3 \times 4$
lattices to find the position of the phase transitions, and 
$12^2 \times 24 \times 4$ lattices to study the transition in detail (the
extension of this lattice in the z direction is to allow the measurement of
hadron screening masses).

\begin{figure}[htb]
\vspace{-1.4in}
\epsfxsize=2.75in
\epsffile{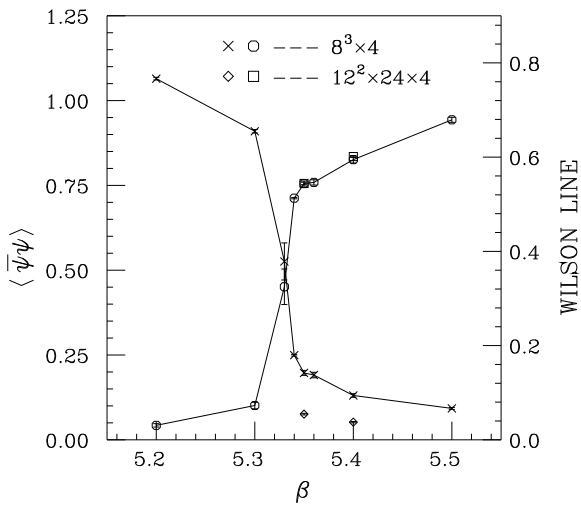}
\vspace{-0.4in}
\label{fig:wil-psi-g10}
\caption{Wilson line and $\langle\bar{\psi}\psi\rangle$ as functions of $\beta$
         for $\gamma=10$}  
\vspace{-0.3in}
\end{figure}

At $\gamma = 2.5$ the deconfinement transition occurs between $\beta=5.5$ and
$\beta=5.6$ while the chiral transition is at $\beta \sim 10$. At $\gamma = 5$
the transitions still appear separated, but clearly the chiral condensate feels
the deconfinement transition. By $\gamma = 10$ the 2 transitions are close
enough together that it is no longer clear whether they are distinct -- see
figure~\ref{fig:wil-psi-g10}.
The way we measure $\langle\bar{\psi}\psi\rangle$, it does not vanish in the
symmetric phase, but should instead behave as $1/\sqrt{V}$. We have tested this
at $\gamma=10$, $\beta=5.4$ and $\beta=5.35$, and find it to be true within
errors which indicates that both these $\beta$'s lie in the chirally symmetric
-- deconfined -- phase. (The phase transition occurs at $\beta \approx 5.33$). 

\section{ZERO TEMPERATURE $\chi$QCD}

We are simulating $\chi$QCD with $N_f=2$ and $m_q=0$ at $\gamma=10$ and 
$\beta=5.4$ on an $8^3 \times 24$ lattice to study zero temperature physics
and in particular hadron spectroscopy. 

\begin{figure}[htb]
\vspace{-1.5in}
\epsfxsize=3.0in
\epsffile{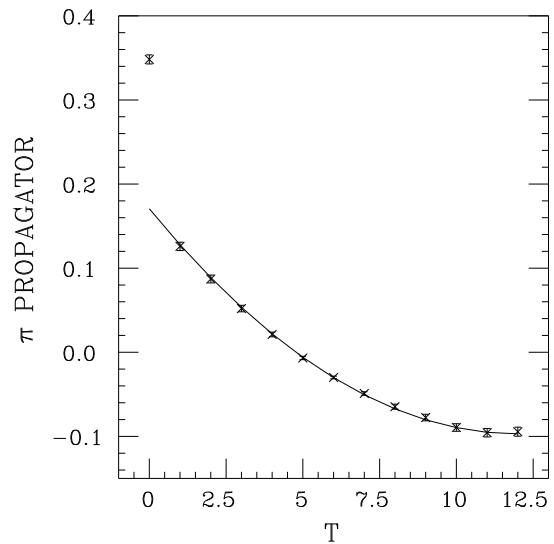}
\vspace{-0.25in}
\label{fig:pi}
\caption{The propagator for the Goldstone pion. The fit is to a massless scalar
         lattice propagator.}
\vspace{-0.25in}
\end{figure}

We have measured the propagators of the $\sigma$ and $\pi$ auxiliary fields on
2500 configurations separated by 2 time units. The first 500 were discarded
leaving 2000 for analysis. These were binned into 200 bins of 20 time units in
length for error analysis and fitting. The pion propagator was fitted to the
massless scalar lattice propagator (see figure~\ref{fig:pi}). The fit, over 
time interval 1--12, had confidence level 78\% showing that it is indeed a 
Goldstone pion. 
The $\sigma$ propagator is noisy, as expected. The best fit we were able
to obtain gave a mass of 1.12(23). Current statistics preclude determining
whether there is any significant contribution from the 2-pion cut. 

So far we have stored 250 configurations spaced by 20 time units for 
calculating the hadron spectrum. We plan first to calculate the local meson
and baryon spectra. For the mesons, this will include the normal, chirality 1
mesons (quarks having chirality $\pm\frac{1}{2}$) formed from linear
combinations of
\begin{equation}
      {\rm Tr}[G(\pi)G^{\dag}(-\pi)]
\end{equation}
and the chirality 0 mesons formed from linear combinations of
\begin{equation}
      {\rm Tr}[G(\pi)G^{\dag}(\pi)]
\end{equation}
where $G$ is the quark propagator. For the normal meson propagators, the
$\sigma$ and $\pi$ have disconnected contributions which we will also need to
calculate, while the $\rho$ and $a_1$ do not. For the chirality 0 mesons there
are no disconnected contributions. 

Working at $m_q=0$ we will need a careful finite size analysis of the hadron 
spectrum, since hadron sizes are determined by $m_\pi$.

\section{$\chi$QCD AT ZERO GAUGE COUPLING}

The $\beta \rightarrow \infty$ limit of $\chi$QCD with $N_f=2$ is the 4-d
Nambu--Jona-Lasinio---Gross--Neveu model with $N_f=6$. We have studied this
theory on an $8^4$ lattice to find its chiral transition, which enables us to
identify its strong and weak coupling domains. We find the transition $\gamma$
to be $\sim 1.7$. We have also studied its finite temperature behaviour on an
$8^3 \times 4$ lattice, where the transition occurs at $\gamma \sim 1.5$. 

\section{COMMENTS ON $\chi$QCD AT FINITE BARYON NUMBER DENSITY}

Quenched lattice QCD has problems at finite baryon number density. If $\mu$ is
the quark number chemical potential, one would expect QCD to have a phase
transition at $\mu \approx m_N/3$. Instead, at $m_q=0$, quenched QCD appears to
have a transition at $\mu=0$ \cite{barbour}.

The nearest thing to a quenched version of massless $\chi$QCD we know is
quenched QCD with fermions interacting with a mean field approximation to the
chiral fields in which $\sigma$ is replaced with a constant and the $\pi$ with
zero. This is just massive quenched QCD ($mass=<\sigma>$) at finite chemical
potential, which has a transition at $\mu \approx m_\pi(<\sigma>)/2$ which is
no longer zero. Since we expect that $m_\pi(<\sigma>)/2 < m_N(<\sigma>)/3$
we have not completely solved the problem. However, for moderate $<\sigma>$,
the behaviour is more physical than for conventional quenched QCD --- the
unphysical region from $m_\pi(<\sigma>)/2$ to $m_N(<\sigma>)/3$ is small.

\section{\bf SUMMARY AND CONCLUSIONS}

$\chi$QCD enables simulations at physical u and d quark masses with at least an
order of magnitude saving in CPU time. It also enables simulations with zero
quark masses which is important for determining the equation of state.
A renormalization group analysis will be needed to continue to the continuum
limit.

After finishing this first round of simulations and hadron mass/screening mass
measurements at zero and finite temperature, we will move to larger lattices,
and also perform some simulations at non-zero quark mass. The next stage will
involve using a more physical chiral group ($SU(2) \times SU(2)$??). 

Auxiliary fields and constituent quark masses give promise of an effective
Lagrangian interpretation of $\chi$QCD. These effective quark masses could
(partially) resolve the problems at finite baryon number density.

Similar theories have been considered by Brower, Orginos, Shen and Tan 
\cite{brower}. Theories with 4-fermion and gauge interactions have a long
history; see for example \cite{kmy}.

\end{document}